\documentclass[aps,prl,twocolumn,showpacs,amsmath]{revtex4}

\usepackage{amsbsy}
\usepackage{amssymb}
\usepackage[dvips]{graphicx}
\usepackage{epsfig}         
\usepackage{psfig}
\usepackage{graphicx,epsf,psfrag}

\newtheorem{theorem}{Theorem}

\DeclareMathOperator{\sign}{\mathrm{sign}}

\newcommand{\ind}{\mathbf{1}}

\newcommand{\R}{\mathbb{R}}
\newcommand{\Z}{\mathbb{Z}}

\newcommand{\cL}{{\ensuremath{\mathcal L}} }

\newcommand{\cD}{{\ensuremath{\mathcal D}} }

\newcommand{\bP}{{\ensuremath{\mathbf P}} }
\newcommand{\bE}{{\ensuremath{\mathbf E}} }

\DeclareMathSymbol{\leqslant}{\mathalpha}{AMSa}{"36} 
\DeclareMathSymbol{\geqslant}{\mathalpha}{AMSa}{"3E} 
\DeclareMathSymbol{\eset}{\mathalpha}{AMSb}{"3F}     

\newcommand{\ga}{\alpha}
\newcommand{\gb}{\beta}

\newcommand{\go}{\omega}

\makeatletter
\def\captionfont@{\footnotesize}
\def\captionheadfont@{\scshape}

\newcommand{\tf}{\textsc{f}}

\begin{document}

\title{Smoothing of Depinning Transitions for Directed Polymers with Quenched Disorder}

\author{G. Giacomin}
\affiliation{Laboratoire de Probabilit{\'e}s de P 6\ \& 7 (CNRS U.M.R. 7599)
  and  Universit{\'e} Paris 7 -- Denis Diderot,
U.F.R.                Math\'ematiques, Case 7012,
                2 place Jussieu, 75251 Paris cedex 05, France}

\author{F. L. Toninelli}
\affiliation{Laboratoire de Physique, ENS Lyon (CNRS U.M.R. 5672), 46 All\'ee d'Italie, 
69364 Lyon cedex 07, France
}


\begin{abstract}
We consider disordered models of 
pinning of directed  polymers on a defect line,
including $(1+1)$--dimensional 
interface wetting models, disordered Poland--Scheraga models of DNA denaturation
and other $(1+d)$--dimensional polymers in interaction with columnar defects.
We consider also random copolymers at a selective interface.
These models are known
to have a (de)pinning transition at some critical line
in the phase diagram. In this work we prove that,
as soon as disorder is present, the transition is at least of second order:
 the free energy is differentiable at the
critical line, and the order parameter ({\em contact fraction}) vanishes continuously at the transition. 
On the other
hand, it is known that the corresponding   non--disordered
models can have
a first order (de)pinning transition, with 
a jump in the order parameter. Our results confirm predictions based on the  {\em Harris criterion}.
\end{abstract}


\maketitle

Quenched disorder, even in  arbitrarily small concentration, is expected to modify qualitatively the 
critical behavior of pure systems in many situations. 
For instance, for Ising spin systems in dimension $d\le 2$ and
for systems with continuous symmetry and  $d\le 4$ it was proven \cite{cf:AW}, 
via a rigorous version of the Imry--Ma argument \cite{cf:Imry-Ma}, that 
randomness in the field conjugated to the order parameter smooths first order phase transitions.
An analogous result was proven \cite{cf:BK} for SOS effective interface models in $(2+1)$ dimensions.

In this Letter, we report on a similar phenomenon in a very different context, i.e., 
for $(1+d)$--dimensional directed polymer models
interacting with a {\em defect line}, and for disordered copolymers at selective interfaces. Both systems
are known to undergo a {\em (de)pinning phase transition}. Such models have natural applications, e.g., 
to biopolymers 
\cite{cf:GHLO,cf:CH,cf:SW}, to
pinning/wetting problems \cite{cf:Derrida,cf:Fogacs},
to the problem of depinning of flux lines from columnar defects in type--II superconductors
\cite{cf:Nelson} and to 
inhomogeneous surface growth equations \cite{cf:KL}, and attracted much attention lately, both in the theoretical physics and in the 
mathematical literature. 
The pure (i.e., non--random) 
models present a variety of critical behaviors, ranging from 
first  to infinite order phase transitions. On the contrary we prove that, as soon as disorder is present,
the transition is always smooth.
In a way, it is remarkable that one can prove such a general result on the nature of the transition, when the knowledge
about  the (de)pinning mechanism itself and about the location of the critical curve is still quite poor.
Our result has interesting implications, in particular, on the nature of the denaturation transition 
for inhomogeneous Poland--Scheraga models of DNA.
\paragraph*{Pinning/wetting models.} We consider general models of directed polymers in interaction with a defect line.
The seemingly abstract setting will be clarified below by
some physically relevant examples.
Polymer configurations are sequences $S=\{S_n\}_{n=0,1,\ldots}$ with values in a set $\Sigma$ 
which contains a  specific point 0 (the origin). We set $S_0=0$.
The {\sl free polymer}, in absence of interaction with the defect line, 
is described by a homogeneous Markov chain on $\Sigma$, with law $\bP$. Our only
assumption  on $\bP$ is the following: let $0=:\tau_0<\tau_1<\ldots$ be the return times to $0$ of $S$  (of 
course, $\tau_i-\tau_{i-1}$ are independent identically distributed (IID) random variables). We require
\begin{equation}
  \label{eq:K}
  K(n)\equiv \bP(\tau_i-\tau_{i-1}=n)\sim n^{-\alpha}, \;n\to \infty
\end{equation}
for some $1\le \alpha<+\infty$. Logarithmic corrections to the power decay \eqref{eq:K} are allowed (and actually
required for $\alpha=1$, to make $K(\cdot)$ summable). 
Note that the first return time $\tau_1$ has infinite mean as soon as $\alpha<2$. 
$S$ may be transient, i.e. $\bP(\tau_1=\infty)>0$. 
As an  
 example, if $S$ is the simple random walk on $\Sigma=\Z^d$, then $\alpha=3/2$ for $d=1$ and
$\alpha=d/2$ for $d\ge 2$. In this case, $S$ is transient as soon as $d\ge3$.
On the  line $S\equiv 0$ are placed quenched IID
random charges $\{\omega_n\}_{n=1,2,\ldots}$ of mean zero and variance one. 
For $\beta\ge0$ and $h\in \R$, the Boltzmann distribution for the polymer of length $N$ is 
\begin{equation}
  \label{eq:gibbs}
  {\bf P}^{\beta,h}_{N,\omega}(S)= {\bf P}(S)\frac{e^{H^{\beta,h}_{N,\omega}(S)}}{Z^{\beta,h}_{N,\omega}}
{\bf 1}_{\{S_N=0\}},
\end{equation}
where  $H^{\beta,h}_{N,\omega}(S)=\sum_{n=1}^N(\beta\omega_n-h){\bf 1}_{\{S_n=0\}}$ and of course
\begin{eqnarray}
  \label{eq:Z}
 Z^{\beta,h}_{N,\omega}=\bE\left( e^{H^{\beta,h}_{N,\omega}(S)}{\bf 1}_{\{S_N=0\}}\right),
\end{eqnarray}
 $\bE$ denoting average with respect to $\bP$.
Note that the polymer--defect interaction takes place only at the contact points, and that a contact at $n$ with
 $(\beta\omega_n-h)>0$ is
  energetically favored. On the other hand, 
polymer configurations which wander away from the line are much more numerous, and therefore entropically favored,
with respect to  those which stay close to it.
The main question is whether   the interaction is enough
to {\em pin} the polymer to the line. 
The infinite volume free energy of the model,
\begin{equation}
  \label{eq:F}
  \tf(\beta,h)=\lim_{N\to\infty}\frac1N \log Z^{\beta,h}_{N,\omega},
\end{equation}
is self--averaging \cite{cf:BdH,cf:MW}. 
Moreover, $\tf(\beta,h)\ge0$, as is seen restricting
the partition function to the configurations which do not touch the defect line between $1$ and $N-1$: these
paths have zero energy, and their entropy is not extensive, in view of \eqref{eq:K}.
One then defines the pinned (or localized) region as
$$
\mathcal L=\{(\beta,h): \tf(\beta,h)>0\}
$$ 
and the depinned (delocalized) region as 
$$
\mathcal D=\{(\beta,h): \tf(\beta,h)=0\}.
$$ 
The denominations pinned/depinned actually correspond to
the typical polymer behavior.  In $\mathcal L$ the polymer stays close to the defect line and touches it
$O(N)$ times before the endpoint 
(various refinements of this statement are proved e.g. in \cite{cf:Sinai} for a related model, the 
copolymer introduced below, and more recently in 
\cite{cf:GTloc} in a more general context). On the other hand, in $\mathcal D$ the number of contacts 
is at most
$O(\log N)$ \cite{cf:GT}. The regions $\mathcal D$ and $\mathcal L$ are separated by the {\em critical line}
$h_c(\beta)$, so that $\mathcal D=\{(\beta,h): h\ge h_c(\beta)\}$.

The above model has a wide range of applications, and a vast literature is dedicated to it. 
Let us mention two particularly interesting examples:

$\bullet$ {\em $(1+1)$--dimensional wetting of a disordered substrate}
\cite{cf:Fogacs,cf:Derrida,cf:AS}. Here, $\Sigma=\Z^+$ and $\alpha=3/2$. 
The defect line represents a wall with impurities,
and $S$ the interface between two coexisting phases (say, liquid below the interface and vapor above). 
$h<0$ means that the underlying homogeneous substrate repels the liquid phase, and vice versa for $h>0$.
$\cL$ corresponds then to the 
{\em dry phase} (microscopic liquid layer at the wall) and  $\cD$ to the {\em wet phase} (macroscopic layer).
One of the most debated  (and still unsettled) issues is whether or not  the critical line coincides  with that of the 
(exactly solvable) {\em 
annealed model}, where disorder is  averaged in the partition function on the same footing
as $S$. 

$\bullet$
{\em Poland--Scheraga (PS) models of DNA denaturation} \cite{cf:KMP,cf:CH}. In this case $\Sigma=\Z^+$,
and $S_n$ represents the relative distance between two DNA strands in correspondence of the $n^{th}$ base pair:
 $S_n=0$ if the pair is bound, $S_n>0$ if the bond is broken. Therefore,
$\cL$ (resp. $\cD$) represents the bound (resp. denaturated) phase.
Modeling $S$ as a simple random walk is known not to be physically realistic, and a phenomenological value 
$\alpha >2$ ({\em loop exponent}), which keeps 
into account the self--avoidance of the two strands, has been
 proposed \cite{cf:KMP}. Therefore, 
the transition is first order in the pure case, cf. \eqref{eq:a>2} below. 
Of course, real DNA is intrinsically non--homogeneous and
one resorts very naturally to disordered models like \eqref{eq:gibbs},
 although the IID assumption on $\omega$ is very questionable in this case.


\paragraph*{Smoothing of the transition.} The order parameter associated to the 
transition 
is the {\em contact fraction},
$$
f_N=N^{-1}\bE^{\beta,h}_{N,\go}\left(\#\{1\le n\le N: {S_n=0}\}\right)
$$
In the pure case ($\beta=0$), critical point and critical behavior can be computed explicitly, see
e.g. 
 \cite{cf:AS,cf:GTsmooth}. 
The critical point is $h_c(0)=\log(1-\bP(\tau_1=\infty))\le 0$ (notice that $h_c(0)<0$ iff $S$
is transient). As for the nature of the transition, one distinguishes two cases: it is
of first order (the contact fraction is discontinuous in the infinite volume limit) if $\sum_{n\ge1}n K(n)<+\infty$,
and of higher order if $\sum_{n\ge1}n K(n)=+\infty$.
In particular, if $\delta\ge 0$ then
\begin{equation}
  \label{eq:a>2}
  \tf(0,h_c(0)-\delta)\sim const\times \delta\;\;\; \text{for}\;\;\; \alpha>2,
\end{equation}
while
\begin{equation}
  \label{eq:a<2}
  \tf(0,h_c(0)-\delta)\sim const\times \delta^{1/(\alpha-1)}\;\;\; \text{for}\;\;\; 1\le\alpha<2
\end{equation}
modulo possible logarithmic corrections.
For $\alpha=1$, the transition is  of infinite  order.

The main result of this Letter is that, as soon as disorder is present ($\beta>0$), the transition is 
always smooth:
\begin{theorem}
\label{th:risultato}
For every $\beta>0$ there exists $0<c(\beta)<+\infty$ such that, for every $1\le \alpha<+\infty$ and $\delta\ge0$,
  \begin{equation}
    \label{eq:risultato}
    \tf(\beta,h_c(\beta)-\delta)\le \alpha c(\beta)\delta^2.
  \end{equation}
\end{theorem}
Notice that, since $ \tf(\beta,h)\ge 0$, \eqref{eq:risultato} is really an estimate 
on the regularity of the transition, an issue debated for example in the
context of the disordered PS model
 \cite{cf:GM2,cf:C,cf:CH}. In particular, \eqref{eq:risultato}
  shows that the order of the transition is at least two, i.e.,
  the fraction of bound 
base pairs vanishes continuously approaching $h_c(\beta)$, 
in contrast with the conclusions of some numerical studies  \cite{cf:GM2,cf:GM3}.
By convexity, self--averaging of $\tf$ implies self--averaging of the contact fraction,
whenever $\partial_h \tf(\beta,h)$ exists.
Theorem \ref{th:risultato} in particular excludes the possibility of
non--selfaveraging
behavior of the contact fraction at the critical point, which was claimed in \cite{cf:GM2,cf:GM3}.
Another interesting consequence of Theorem \ref{th:risultato}
 is an upper bound on the number of pinned sites in a small window around the the critical point,
for finite $N$: indeed, one can show \cite{cf:Tnota}
that,
if $\beta>0$ and $|h-h_c(\beta)|\le const\times N^{-1/3}$,
the probability that $f_N\gg N^{-1/3} $ vanishes for $N\to\infty$.
Note that, comparing  \eqref{eq:a<2} and \eqref{eq:risultato}, 
our result confirms Harris' criterion \cite{cf:Harris} which, translated into the present context, 
predicts that disorder  is relevant and changes the nature of the transition as soon as $\alpha>3/2$
(it also predicts that the critical behavior does not  change if $\alpha<3/2$, which is compatible
with  \eqref{eq:risultato}.) 
For previous rigorous work connected to the Harris criterion and to critical
exponent inequalities for random systems, cf. \cite{cf:CCFS}.

As a last remark, note that Theorem \ref{th:risultato} is reminiscent of the Aizenman--Wehr result \cite{cf:AW} 
about 
smoothing of first order phase transitions via quenched disorder in $2$d spin systems (in particular, the 
Random Field Ising Model). However,  the analogy is rather superficial and 
  very different physical mechanisms are involved in the two cases. Indeed, \cite{cf:AW} is based
on  a comparison between two competing effects: on one hand the ordering effect 
of boundary conditions, on the other the effect of random field fluctuations in the bulk.
In our case, instead,  boundary conditions
play no role at all (the endpoint $S_N$ is pinned to $0$, cf. \eqref{eq:gibbs}).
Our method consists rather in selecting polymer configurations that 
visit rare but favorable 
regions with atypical disorder, and in giving
{\em Large Deviation Estimates} on the number of such regions. 
This approach was partly inspired by \cite{cf:BG}, where a similar {\em path selection 
method} was used to obtain rigorous lower bounds on $\tf(\beta,h)$ for the copolymer model.

\begin{figure}[h]
\begin{center}~
\leavevmode
\epsfxsize =9 cm
\psfragscanon
\psfrag{0}[c][l]{$0$}
\psfrag{n}[c][l]{ $n$}
\psfrag{b1}[c][]{ $B_{0}$}
\psfrag{b2}[c][]{ $B_{1}$}
\psfrag{b3}[c][]{ $B_{2}$}
\psfrag{b4}[c][]{ $B_{3}$}
\psfrag{z}[c][]{ $z$}
\psfrag{b5}[c][c]{ $B_{9}$}
\psfrag{dots}[c][c]{ $\dots$}
\psfrag{L0}[c][c]{ $L_0$}
\psfrag{L1}[c][c]{ $L_1$}
\psfrag{y}[c][c]{ $y$}
\psfrag{x}[c][c]{ $x$}
\psfrag{N}[c][l]{ $N$}
\psfrag{Sn}[c][l]{$S_n$}
\epsfbox{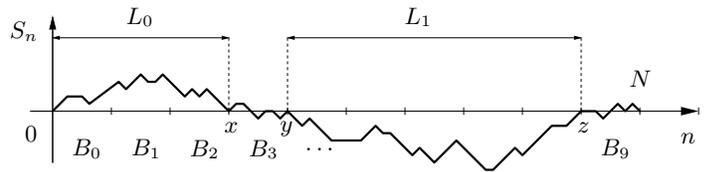}
\end{center}
\caption{\label{fig:figura} A typical trajectory in $\mathcal S_\go$. 
Here $k=10$, $\ell=8$  and $\mathcal I(\omega)=\{3,9\}$. Note that $S_n\ne 0$ for $n$ in   
 $B_j$ with $j \not\in  \mathcal I(\omega)$ ({\em bad blocks}), except at the boundary with a block $B_j$ with 
$j \in \mathcal I(\omega)$.
 On the other hand, inside $B_{j}$,  $j \in \mathcal I(\omega)$ ({\em good blocks}), the walk moves without constraints.
The {\sl excursions} $L_0, L_1,\ldots$ are typically of length $\ell \exp(\ell \delta^2/2)$.
The polymer is pinned to zero at steps $\{0,x,y,z,N\}$, so that $Z_{N,\omega}^{\beta,h}$ factorizes into $4$ terms.}
\end{figure}

\paragraph*{A Large Deviations approach.}
Theorem \ref{th:risultato} 
is proven in full detail in Ref. \cite{cf:GTsmooth}, under some technical assumptions on the 
law of $\omega$: 
the result holds in  particular if $\omega_n$ is bounded or if 
it is Gaussian.  
Here, we present an intuitive argument which clarifies the heart of the method. 
Assume for simplicity  a Gaussian  distribution for the disorder, $\omega_n\sim\mathcal N(0,1)$. Let  $1\ll\ell\ll N$
and divide the system into $k=N/\ell$ blocks $B_0,\ldots,B_{k-1}$ of length $\ell$. For a given disorder realization, 
select the {\em good blocks} where the sum of the charges is approximately $\delta\ell$, i.e., let 
$$
\mathcal I(\go)=\left\{0\le j\le k-1: \sum_{n=\ell j+1}^{\ell (j+1)} \omega_n\sim \ell \delta\right\}.
$$
By elementary large deviations considerations, one realizes that 
there are typically $M_{typ}=(N/\ell) e^{-\ell \delta^2/2}$
good blocks, two successive good 
blocks being separated by a typical distance $d_{typ}=\ell\, e^{+\ell \delta^2/2}$.
Next, select all those configurations of $S$ that {\em touch} $0$ at the endpoints of the good blocks $B_j, 
j\in \mathcal I(\omega)$ and that {\em do not touch} $0$ inside the bad blocks $B_j, j\notin \mathcal I(\omega)$ 
 (cf. Fig. \ref{fig:figura}), and call $\mathcal S_\omega$ the collection of such configurations.
Of course, one obtains a lower bound on the free energy by restricting the partition sum to the selected 
configurations, i.e.,
\begin{eqnarray}
  \label{eq:selection}
  \frac1N \log Z^{\beta,h}_{N,\omega}\ge \frac1N \log 
\bE\left( e^{H^{\beta,h}_{N,\omega}(S)}{\bf 1}_{\{S\in \mathcal S_\omega\}}
{\bf 1}_{\{S_N=0\}}\right).
\end{eqnarray}
Thanks to the Markov property of $\bP$, the r.h.s. of \eqref{eq:selection} factorizes into a product of terms, one
for each good block and one for each excursion corresponding to a group of adjacent bad blocks (cf. Fig. \ref{fig:figura}).
Note that conditioning  $\ell$ independent Gaussian variables to  have sum $\delta \ell $ is equivalent, for $\ell$
large, to shifting the mean of each variable from $0$ to $\delta$, while 
keeping their variance at $1$. Therefore, in each of the good blocks 
the polymer effectively has thermodynamical parameters $(\beta',h')=(\beta,h-\delta \beta)$.
Also, note that each of the long excursions between two good blocks
entails an entropic loss $\log K(d_{typ})\sim -\alpha \ell \delta^2/2+O(\log \ell)$, cf. \eqref{eq:K}. 

Now, take the system {\em at} the critical point, $h=h_c(\beta)$, and let $N\to\infty$ in \eqref{eq:selection}.
By the law of large numbers, the free energy contribution of good blocks converges to their {\sl density},
$\rho=\ell^{-1}\, e^{-\ell \delta^2/2}$, times
the {\sl average} 
contribution of each of them, which is $\ell\, [\tf(\beta,h_c(\beta)-\delta \beta)+o(1)]$ for $\ell$ large
(here and below the {\sl error term} $o(1)$ denotes a non--random quantity that vanishes as $\ell \to \infty$).
Similarly, the contribution of excursions converges to $\rho$ times 
$\ell(-\alpha \delta^2/2+o(1)).$
In formulas,  \eqref{eq:selection} implies
\begin{eqnarray}
\nonumber
  \label{eq:LB}
  0=\tf(\beta,h_c(\beta))
\ge  e^{-\frac{\ell \delta^2}2} 
\left(\tf(\beta,h_c(\beta)-\delta \beta)
-  \frac{\ga \delta^2}2+o(1)\right).
\end{eqnarray}
Therefore $\tf(\beta,h_c(\beta)- \delta \beta)\le
\alpha  {\delta^2}/2+o(1)$ for every finite $\ell$. Since 
$\ell$ is arbitrary, we obtain
 \eqref{eq:risultato}.

\paragraph*{Copolymers at selective interfaces.} Consider a polymer chain  close to the interface between two
solvents A and B, and assume that some of the monomers have a 
larger affinity with A and  others 
with B. If the monomers are placed inhomogenously along the chain, the energetically most
favored configurations will stick close to the interface. The competition with entropic effects produces also in 
this case a non--trivial (de)localization transition at the interface.
This model was introduced in the physical literature 
\cite{cf:GHLO,cf:Monthus} and has attracted a lot of attention in the mathematical one,
cf. e.g. \cite{cf:Sinai,cf:BdH}.
The system, although physically $3$--dimensional, can be reduced to a $(1+1)$--dimensional one if  
 self--avoidance of the polymer  is neglected \cite{cf:GHLO}. Its Boltzmann distribution 
can be expressed, in analogy with \eqref{eq:gibbs}, as
\begin{equation}
  \label{eq:gibbs2}
  {\bf \widehat P}^{\beta,h}_{N,\omega}(S)= {\bf P}(S)\frac{e^{\frac12
\sum_{n=1}^N(\beta\omega_n+h)\sign(S_n)}}{\widehat Z^{\beta,h}_{N,\omega}}
{\bf 1}_{\{S_N=0\}},
\end{equation}
with the convention $\sign(0)=+1$.
Here, the natural setting is to take $S$ as a symmetric Markov chain on 
$\Sigma=\Z$, with  increments $S_n-S_{n-1}\in\{-1,0,+1\}$. By symmetry, one can take $h\ge0$. 
The random variables $(\gb\go_n+h)$ express the affinity of the $n^{th}$ monomer with A,
and $h$ is a measure of the asymmetry of the chain (if $h>0$ there is typically
a fraction $>1/2$ of 
monomers which prefer A ({\sl favorable solvent})). 
In the literature, the only  case considered is that of symmetric random 
walks with IID  increments $S_n-S_{n-1}$, 
which implies $\alpha=3/2$, but in our approach this restriction is not required and our 
analysis covers more general Markov processes.

Again, one introduces the free energy and, noting that in this case
$\widehat \tf(\beta,h)\ge h/2$ (see e.g. \cite{cf:BdH}), one defines localized and delocalized regions 
$\cL,\cD$ according to whether 
strict inequality holds or not. Replica methods \cite{cf:TM} 
and real--space renormalization group arguments \cite{cf:Monthus} were used to attack the model, and
 rigorous bounds are known for $\widehat \tf(\beta,h)$ and 
for the critical curve $h_c(\beta)$ separating $\cL$ and
$\cD$ \cite{cf:BdH,cf:BG}. 
Interestingly, recent numerical simulations plus probabilistic arguments
indicate that none of the known bounds is optimal \cite{cf:CGG}, 
which means that the (de)localization mechanism is still poorly understood.

While the physics of pinning/wetting models and of copolymers are rather different,  the
approach  we present here is rather robust and works equally well for 
the two problems. Indeed,
also for the copolymer model we can prove smoothness of the (de)localization 
transition for all $\beta>0$ and $1\le\alpha<\infty$: Theorem \ref{th:risultato} still holds, 
with $\tf(\beta,h)$ replaced by $\widehat \tf(\beta,h)- h/2$ 
\cite{cf:GTsmooth}, so that the transition is at least second order 
in view of $\widehat \tf(\beta,h)-h/2\ge0$. 
Here, we give just an idea of how the heuristics above must be modified to obtain the result in this case.
The main point is that \eqref{eq:gibbs2} can be rewritten as
\begin{equation}
  \label{eq:gibbs3}
  {\bf \widehat P}^{\beta,h}_{N,\omega}(S)\propto {\bf P}(S)e^{-\sum_{n=1}^N(\beta\omega_n+h)\Delta_n}{\bf 1}_{\{S_N=0\}},
\end{equation}
where 
$\Delta_n=0$ if $\sign(S_n)=+1$ and $1$ otherwise. In this form, the analogy with \eqref{eq:gibbs} becomes more
evident, the 
role of $\ind_{\{S_n=0\}}$ being played by $\Delta_n$. 
One can again divide the system into blocks and select good ones where the sum of the charges
is atypically 
large. However, when the selection of trajectories is performed as in \eqref{eq:selection}, an 
extra condition has to be met: the selected trajectories, $S\in \mathcal S_\go$, 
must satisfy $\Delta_n=0$ for $n$ in a {\em bad block} (which means $\sign(S_n)=+1$, and not just $S_n\ne 0$).
 Apart from that, the argument is  identical as for pinning models.

Finally, note that for the copolymer the order parameter is no longer
 the contact fraction $f_N$, but rather
$\widehat f_N=N^{-1}\widehat\bE_{N,\go}(\#\{1\le n\le N: S_n<0\})$, 
i.e., the fraction of monomers in the (unfavorable) solvent B, which
vanishes continuously at the transition, in view of our result. 
Again, one can prove finite--size upper bounds of order $N^{-1/3}$ for $\widehat f_N$, around
the critical point \cite{cf:Tnota}.

\paragraph*{Conclusions.} We have proved that an arbitrarily small amount of disorder is enough 
to smooth the (de)pinning transition in  directed (co)polymer models. In particular, 
the transition is always at least of second order, even when it is
discontinuous in the corresponding pure models. Moreover, we have given
 finite $N$ estimates on the order parameter 
{\em at} the critical point. In some literature, it is  conjectured 
that the transition is actually of order higher than two (possibly infinite) in some situations: 
in particular, for the copolymer and pinning models 
with $\alpha=3/2$ \cite{cf:Monthus,cf:TM,cf:TangChate}. 
Our result leaves this possibility open.

We  thank Bernard Derrida, Thomas Garel, Massimiliano Gubinelli, C\'ecile Monthus and David Mukamel for fruitful discussions. Work partially supported by GIP--ANR project 
 JC05\_42461
({\sl POLINTBIO}).

\end{document}